\documentclass{chjaa}                  

\usepackage{graphicx,times}             
\input{epsf.sty}                        
\input{psfig.sty}                       

\usepackage[authoryear,sort&compress]{natbib}
\usepackage{amsmath,amssymb}            

\begin{document}

   \title{Testing Lorentz Violation Using Propagating UHECRs
}

   \volnopage{Vol.0 (200x) No.0, 000--000}      
   \setcounter{page}{1}           

   \author{Cong-Xin Qiu
      \inst{}\mailto{congxin.qiu@gmail.com}
   \and Zi-Gao Dai
      \inst{}
      }

   \institute{Department of Astronomy, Nanjing University, Nanjing 210093, P. R. China
             \email{congxin.qiu@gmail.com}
          }

   \date{Received~~2001 month day; accepted~~2001~~month day}

   \abstract{
Lorentz invariant violation (LIV) test is very important to study in
the new physics. All the known astrophysical constraints either have
a very small examinable parameter space, or are only suitable for
some special theoretical models. Here we suggest that it is possible
to detect the time-delay of ultra-high-energy cosmic-rays (UHECRs)
directly. We discuss some difficulties in our method, including the
intergalactic magnetic fields. It seems that none of them are
crucial, hence this method could give a larger examinable parameter
space and a stronger constraint on LIV.
   }

   \authorrunning{C.-X. Qiu \& Z.-G. Dai}            
   \titlerunning{To Test Lorentz Violation Using Propagating UHECRs}  

   \maketitle

\section{Introduction}

Lorentz invariant violation (LIV)
test~\citep{196700000_PhysRev.159.1106_Pavlopoulos} is significant
for the applicability of Special Relativity, even if it has few
trustworthy theoretical foundations. The theoretical approaches for
LIV are mainly from new physics, including the Standard-Model
Extension, noncommutative geometry~\citep{200300000_Szabo_physRep},
loop quantum
gravity~\citep{199800000_Rovelli_LoopQuantumGravity_review_at_livingreviewsOrg}
and string
theory~\citep{book_1987_Green_Schwarz_Witten,book_1998_Polchinski}.

The Standard-Model Extension approach~\footnote{\textless
http://www.physics.indiana.edu/\~{}kostelec/faq.html
\textgreater}~\citep{199700000_Colladay,199800000_Colladay} is the
most straightforward one, which introduces LIV as an assumption. LIV
may be caused by spontaneously violating the vacuum solution, if not
by the theory itself. The minimum Standard-Model Extension wishes to
maintain all the conventional desirable properties of the
Standard-Model beside allowing for violations of Lorentz symmetry,
hence it does not have many astrophysical (time-integral-kind)
applications. However, the Standard-Model Extension can actually
induce some kind of birefringence effects for
photons~\citep{200100000_Kostelecky}.

Noncommutative geometry has a lot of phenomenological applications.
However, most of them are based on terrestrial
experiments~\citep{200205040v3_Hinchliffe_phenomenology,200200000_Konopka_NewJournalPhy_ObsLimits_on_QuantumGeometryEffects}.
The derivation of \emph{particle} Lorentz-violating terms from
noncommutative
geometry~\citep{200100000_PhysRevLett.87.141601_Carroll} seems more
natural than other approaches, but unfortunately, it does not have
(at least we don't know how it can have) a beautiful and feasible
way to be approved by time-integral-kind experiments. The
differences are as follows. Some theoretical models result in a
constant space of light (e.g. by $\kappa$-Minkowski
space-time~\citep{200200000_PhysRevD.65.083003_Tamaki}). And,
researchers seem to have different opinions on whether the
Lorentz-violating term $\theta^{\mu\nu}$ depends or not on position,
energy or momentum~\footnote{The kind of experiments we are
interested in work only if $\theta^{\mu\nu}$ depends on energy and
momentum, but is independent or almost independent of position for
us to integral the effect.}.

The loop quantum gravity approach seems the most usable one. The
propagational calculations of
photons~\citep{199900000_PhysRevD.59.124021_Gambini,200200000_PhysRevD.65.103509_Alfaro}
and neutrinos (or other massive spin-$1/2$
fermions)~\citep{200000000_PhysRevLett.84.2318_Alfaro,200200000_PhysRevD.66.124006_Alfaro}
are fulfilled. The propagation speeds in both cases are non-trivial,
with velocity departure linearly depending on particle energy.
Further more, photons have a first order and neutrinos have a second
order birefringence effect. However, although ``foam''
structure~\citep{incollection_1964_Wheeler,197800000_Hawking,200303037v1_Doplicher_quantum_structure_of_spacetime,199800000_PhysRevLett.80.2508_Garay}
is really an intuitive way to understand the nature of quantum
space-time, we have to warn ourselves time and again that loop
quantum gravity theory itself has some theoretical
problems~\citep{199200000_PhysRevLett.69.237_Ashtekar}, ``weave''
states and coarse graining approximation are at most effective
models.

Another leading (and in fact, chronologically the ``first'') root
for the LIV calculations is from Liouville
string~\citep{199207103v2_Ellis_PLB_Liouville_strings}, a
phenomenological model which makes the calculations of propagation
equation in the framework of string theory
possible~\citep{199605211v1___Amelino-Camelia_distance_measurement,200000000_Ellis_GR_and_Gravitation_c_from_string}.
What they can calculate are the so-called ``photons'' which are the
endpoints of open-strings attached to D-branes, and the space-time
foam is described by D-brane fluctuations. The model can result in
Lorentz-violating propagation equation by LIV of string ground
state, although it also has some inconsistencies. As a result, LIV
is stochastic, and the degree of velocity departure is first
ordered. However, there is no evidence to support birefringence,
which is in conflict with loop quantum gravity results.

There are also some other ways to discuss LIV from the theoretical
viewpoint, although some of them are formerly due to the so-called
GZK anomaly, which may in fact be some kind of experimental
errors~\citep{200703099v1,200700000_Cho_science_UHECR_from_galaxies}.
The methods include simply adding tiny (first order or second order
or whatever we want) Lorentz-violating terms to a conventional
Lagrangian (these may be considered as some kind of Standard-Model
Extension) and seeing how they can affect our
observations~\citep{199900000_PhysRevD.59.116008_Coleman,200300000_PhysRevLett.90.211601_Myers},
calculating the geodesic in a topological fluctuated
\emph{classical} general relativity to get some very complicated
results~\citep{199900000_PhysRevD.60.084023_Yu}, deforming the
measure of integration in Feynman graphs (which is equivalent to
inventing a new renormalization skill) to get an effective
LIV~\citep{200500000_PRL_alfaro:221302,200500000_PRD_alfaro:024027},
calculating the graviton induced corrections to Maxwell's
equations~\citep{200100000_PhysRevD.63.084023_Dalvit}; however, the
resultant speed of light correction in the last method is
independent of energy. A recent work by
\cite{200700000_Gogberashvili} deduces the dispersion relation (with
no birefringence effect) from fat brane-world scenario; but the
resultant constraint seems to be too strict to trust that model.

Astrophysical experimental (dis)confirmations of LIV often use far
transient sources emitting high-energy particles. The common sources
are gamma-ray bursts (GRBs) which are cosmological, have very short
durations, and can emit high-energy
photons~\citep{2007041329_notforQG_buthighEnergyPhotons_fromGRB} and
neutrinos~\footnote{There are really a lot of different models for
GRBs to emit ultra-high energy neutrinos, from
\citep{199700000_PhysRevLett.78.2292_Waxman} until now. Nearly all
of the scenarios are $p+p \mbox{\ or\ } p+\gamma \Rightarrow \pi^{+}
\Rightarrow \nu$, but in different environments. See
\cite{200100000_Waxman_NPB_review} for a review.}. The other common
sources are giant $\gamma$-ray flares of active galactic nuclei
(AGNs); however, there have not been suitable models for the shapes
of the time profiles until now. If energy can affect particle speed
by the LIV effect (it's not the same as the effect of particle mass,
which becomes unimportant if the particle is sufficiently
energetic), as some theoretical works predicted, particles emit
simultaneously from the source but with different energies will
exhibit a time-delay when observed. The possible ways include
testing the time-delay of prompt emission photons from
GRBs~\citep{199800000_Amelino-Camelia_nature_393763a0_GRB_test_QG,199912136_Norris_GLAST_GRBs_QG,200000000_Ellis_ApJ_search,200300000_Ellis_AA,200600000_Ellis_AstropartPhy}
and giant $\gamma$-ray flares of
AGNs~\citep{199900000_PhysRevLett.83.2108_Biller,2007082889v1_Mkn501Flare_MAGIC},
the time-delay of neutrinos from
GRBs~\citep{200000000_PhysRevLett.84.2318_Alfaro,200200000_PhysRevD.66.124006_Alfaro,200000000_Bertolami,200300000_PhysRevD.67.073005_Choubey,200700000_Jacob_Piran_nature_phys506},
the polarized photons from
GRBs~\citep{200100000_PhysRevD.64.083007_Gleiser,200300000_Mitrofanov_nature_426139a_polarization,200702006v1_Fan_GRB_polarization__polarimetry_and_QG}
and distant galaxies~\citep{200100000_Kostelecky} which should be
destroyed by
birefringence~\citep{199900000_PhysRevD.59.124021_Gambini,200200000_PhysRevD.65.103509_Alfaro},
the synchrotron radiation from the Crab
nebula~\citep{200200000_PhysRevD.66.081302_Jacobson,200300000_Jacobson_nature01882}
which should not be observed if photons can be both superluminal and
subluminal but electrons can only be
subluminal~\citep{200300000_PhysRevLett.90.211601_Myers}. There are
also a lot of theoretical works to explain the GZK anomaly by
Lorentz-violating
terms~\citep{199900000_PhysRevD.59.116008_Coleman,200000000_PhysRevD.62.053010_Aloisio,200100000_PhysRevD.64.036005_Amelino-Camelia,200300000_PhysRevD.67.083003_Alfaro},
so if the GZK
cutoff~\citep{196600000_PhysRevLett.16.748_G,196600000_ZK} does in
fact exist, the inverse proportion may also give some kind of
constraints.

The purpose of this paper is to suggest a different way to
(dis)confirm the LIV effect; that is, to test directly the
time-delay of ultra-high-energy cosmic-rays (UHECRs) from far away
sources. This method may give a larger examinable parameter space
and a stronger constraint.

\section{Calculation}

\subsection{Naive Time-Delays by the LIV Effect\label{subsection_time_delay}}

One possible way to (dis)confirm LIV is simply to test the
time-delay of UHECRs from far away sources. Because in mainstream
quantum gravity models, the departure of velocities depends on
energy
\emph{linearly}~\citep{199900000_PhysRevD.59.124021_Gambini,200200000_PhysRevD.65.103509_Alfaro,200000000_PhysRevLett.84.2318_Alfaro,200200000_PhysRevD.66.124006_Alfaro,199605211v1___Amelino-Camelia_distance_measurement,200000000_Ellis_GR_and_Gravitation_c_from_string}
in the massless approximation, the time-delay is very sensitive to
ultra-high-energy particles. A naive calculation shows that the
time-delays are really huge. For example, in the standard
cosmological model where $H_{0}$, $\Omega_\mathrm{m}$ and
$\Omega_{\Lambda}$ as the customary cosmological parameters, the
propagation equation and time-delay for a massless particle has the
form
\begin{equation}
    v = c \left( 1 \pm \frac{E}{\xi E_{\mathrm{pl}}} \right)
\end{equation}
and
\begin{equation}
    \Delta t_{\mathrm{QG}} =
        \frac{1}{H_{0}} \int_{0}^{z} \left( \frac{E}{\xi E_{\mathrm{pl}}} \right)
        \frac{(1 + z') d z'}{\sqrt{\Omega_\mathrm{m} (1+z')^3 +
        \Omega_{\Lambda}}} \mbox{,}
    \label{eq_time_delay}
\end{equation}
where $E \ll E_{\mathrm{pl}}$ is the energy of the particle, $z$ is
the redshift of the source, $\xi$ is a free parameter to describe
the degree of violation (which we want to restrict) with assumed
typical value of unity, $c$ is the speed of light, and
$E_{\mathrm{pl}}$ is the Planck energy. To give a straightforward
example, insert $E = 10^{19.8}\,\mathrm{eV} \simeq 6.3 \times
10^{19}\,\mathrm{eV}$ as the GZK threshold energy, $z = 0.1 \simeq
400\,\mathrm{Mpc}$ as a nearby source, and $\xi = 1$ as a typical
dimensionless free parameter, we have
\begin{equation}
    \Delta t_{\mathrm{QG}} \simeq 7\,\mathrm{yr} \mbox{.}
    \label{equation_an_example}
\end{equation}

We choose $z = 0.1 \simeq 400\,\mathrm{Mpc}$ rather than larger
distances to avoid $\Delta t_{\mathrm{QG}}$ to be too large to be
compared with human longevity. In this case, cosmological models are
in fact irrespective, so the situation differs from considering less
energetic but neutral particles (like photons or neutrinos) that
come from more far away sources. Closer sources are also possible
(and maybe even better), because nearly all the time-delay effects
(including the intergalactic magnetic fields, which we discuss in
detail in \S~\ref{subsection_mag_field}) caused by propagation
depend \emph{linearly} on distance, and distance is irrespective
when contrasting which one of the time-delay effects is more
important. Remote sources are only needed when $\Delta
t_{\mathrm{QG}}$ is too small compared with the internal duration of
the events themselves, which is only several seconds for GRBs and
some other transient sources.

When the energy of the UHECR particles exceeds the GZK threshold, it
is less possible for the source to be too far away, because the
particles lose energy by interacting with CMB photons. The main
mechanisms of energy loss on the road are photomeson
production~\citep{196800000_Stecker_PRL} and $e^{+} e^{-}$ pair
production~\citep{197000000_Blumenthal}, with their mean free paths
already being calculated. However, for the UHECR events with energy
larger than the GZK threshold we have \emph{already} observed, their
time-delays by the LIV effect are really interesting, because they
should be more energetic and more sensitive to LIV when just be
emitted.

\begin{figure}[ht]
\includegraphics[height=13cm]{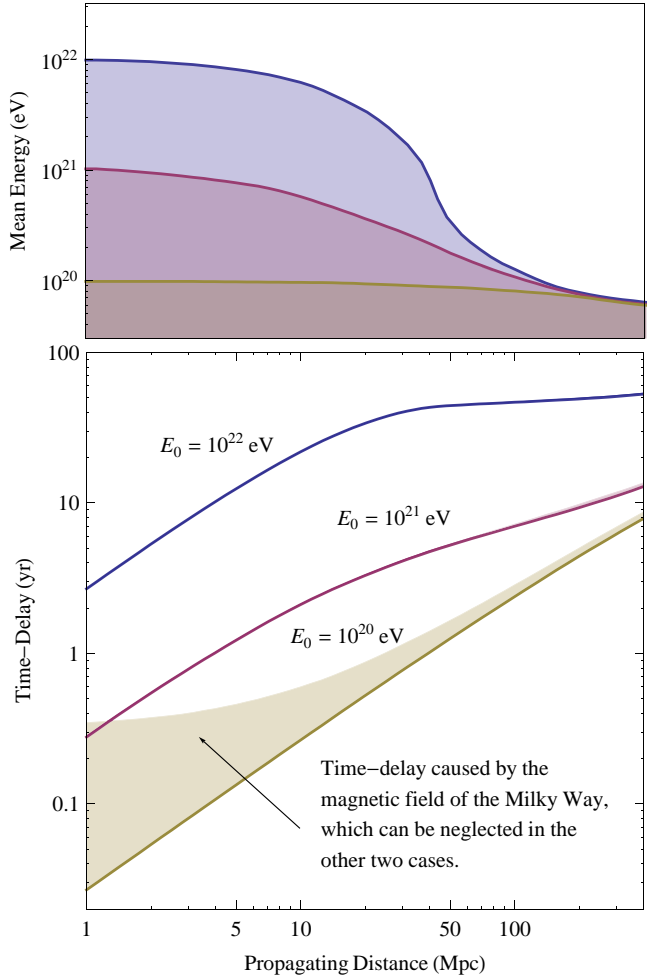}\\
\caption{Top, how particle energies change with propagation
distance, if their initial energies are above the GZK
threshold~\citep{199200000_Cronin}. The initial energies are chosen
to be $10^{20}$, $10^{21}$ and $10^{22}\,\mathrm{eV}$. Bottom, the
total time-delays caused by the LIV effect. The shadow regions are
the time-delays caused by the magnetic field of the Milky Way, which
can be neglectable undoubtedly in the late two cases.}
\label{pic_energy_time_delay}
\end{figure}

However, the calculation of energy loss rate $d E/d x$ (where $x$ is
the propagation distance) is very difficult, although what we have
to face are trivial details of standard quantum field theory and
phase space integrals. Here we simply use the existing numerical
results~\citep{199200000_Cronin,199400000_Aharonian_Cronin} to
proceed our calculations. Time-delays depending on different
propagation distances are shown in
Fig.~(\ref{pic_energy_time_delay}, bottom). We see that although the
time-delays finally tend to the same level as others (because all
their energies converge to the GZK threshold energy after a long way
of propagation), their differences are tremendous when just been
emitted. So, confirming the sources of the UHECR events above the
GZK threshold energy may also be a way to test the LIV effect, even
if the source is really nearby.

We have to emphasize here that the particle energy versus
propagation distance relation in Fig.~(\ref{pic_energy_time_delay},
top) is based on some statistical results with large samples,
because the photomeson interaction is stochastic (see more detailed
discussion in \S~\ref{subsubsection_time-delay_II}). The mean free
paths for UHECRs with energy $E \gtrsim 10^{21}\,\mathrm{eV}$ are
about $10\,\mathrm{Mpc}$, and larger for less energetic
ones~\citep{196800000_Stecker_PRL}. A certain particle can only
suffer from a couple of collisions before being observed, so the LIV
constraint by one single UHECR event has its initial measurement
errors. However, the observed time-delay can at least give an upper
limit for LIV, and we can still improve our result by averaging more
events or using more advanced statistical methods.

\subsection{Problem of the Applicability of the Time-delay Equation}

To investigate the capacity of the method we propose, there are
several problems which have to be considered carefully. The first
problem is whether the naive propagation equation can be used for
UHECRs. Although we are not sure what the compositions of UHECRs
are, air shower data exclude photons or electrons, and the
existence~\citep{200703099v1,200700000_Cho_science_UHECR_from_galaxies}
of the GZK
cutoff~\citep{196600000_PhysRevLett.16.748_G,196600000_ZK} suggests
in fact protons or heavier nuclei. Although we cannot rule out other
possibilities like exotic particles, we will assume in the context
they are protons which are compound and have finite rest mass
(discussions with the assumption that they are heavier nuclei like
Fe are analogous). Mass is not a serious problem at the energy scale
of the GZK threshold~\footnote{Of course, massive and massless
particles are totally different from the viewpoint of quantum field
theory. However, we give up discussing it deeply, because of the
still inconsistent theoretical works.}. For example, in the case of
parameters using above in Eq.~(\ref{equation_an_example}), the
time-delay affected by proton mass is only $5 \times
10^{-6}\,\mathrm{s}$, much shorter than $\Delta t_{\mathrm{QG}}$
affect by the LIV effect. A more serious problem is the complexity
of protons. We all know that proton is made up of three quarks,
therefore using the LIV calculations for \emph{elementary} particles
to calculate it will not be justified. A detailed study of quantum
chromodynamics (QCD) with Lorentz-noninvariant terms is needed;
however, it will certainly be very difficult. We still use the same
propagation equation by some scaling
arguments~\citep{199900000_PhysRevD.59.116008_Coleman}, or by simply
thinking its effect as an overall constant coefficient, just like
that of turning off QCD~\footnote{If we turn off QCD, the
propagation equation can be used to every elementary particle inside
a proton, so the overall effect is only a constant coefficient.}.
For the reason that the LIV confirmation is still qualitatively
rather than quantitatively at present, an overall constant can be
neglected.

\subsection{Problem of the Source}

\subsubsection{Time Bases}

The second problem is how we can choose the time bases, thus compute
the time-delays with a suitable zero point. Because $\Delta t$
should be typically very long, as shown in
Eq.~(\ref{equation_an_example}), it is a serious problem to know
when UHECRs should come if we change their energies (because $\Delta
t_{\mathrm{QG}}$ depends on energy of the particle) or turn off the
LIV effect (as in the classical limit of $E \rightarrow 0$).
Comparisons of events with photons or other low energy massless
particles (which are certainly much less energetic and can be taken
for as in the $E \rightarrow 0$ limit; hence the LIV effect of it
can be neglected) and other UHECRs (with energy different from each
other) from the same source can scale the time-delay. The
precondition is that we have confirmed the source, or that we are
convinced of the fact that different signals come from the same
source.

As the recent powerful
evidence~\citep{200700000_Cho_science_UHECR_from_galaxies} shows,
UHECRs come from some \emph{extragalactic} sources because they are
anisotropic and correlated with the direction of the Super-galactic
plane. In that case, the mainstream models for sources are
AGNs~\citep{book_Ginzburg_CR,198400000_Hillas_ARAA} and
GRBs~\citep{199500000_Waxman_PRL,199506081_Vietri_ApJ1995,200302144_Vietri_ApJ2003,200310667v2_Wick_AstroPartPhy,200210638_Waxman_ApJ2004},
but other sources distributing within the Super-galactic plane are
also possible (if they are related to, e.g. galaxy formation or
stellar formation, which is always true). The main mechanism is
Fermi acceleration but in different environments.

As AGNs are lasting sources, we can hardly know very well
\emph{when} the source emitted the UHECR particle we observed.
However, some recent theoretical work~\citep{2008021074v2} shows
that the UHECR emissions are associated with AGN giant flares, with
typical wait-time about $10^3$ to $10^4$
years~\citep{200200000_Donley_202066web__AGN_gaint_flare}. Because
the duration is much longer than the typical time-delay we gave in
Eq.~(\ref{equation_an_example}), AGNs can be used in this method if
the theoretical work mentioned above is true. GRBs are much better
sources, because nearly all of the mainstream central engine models
(including collapsars, supranovae and mergers of compact objects)
tell us they are transient and burst only one time in their whole
lives. If the emission of UHECRs and the burst itself happen almost
at the same time (which is the most reasonable assumption), we can
scale the time-delay by the observed low energy $\gamma$-rays
because they can hardly be affected by mass, electromagnetic fields
and the LIV effect. Other sources are also possible, if they emit
particles (photons for instance; however, they are not exclusive
choices) other than UHECRs which can be observed by our scientific
equipments.

Are these kind of sources practicable for our purpose? The distances
of the sources mentioned above are all suitable for the constraint
that $\Delta t_{\mathrm{QG}}$ given in
\S~\ref{subsection_time_delay} should not be too large. Short GRBs
are often not too far away from us, and there are already a number
of GRBs with redshift $z \sim 0.1$, including a special one (GRB
980425) with an especially small redshift $z =
0.0085$~\citep{199800000_Galama}. Although the number density of
AGNs decreases quickly when $z < 1$, there are already hundreds of
nearby AGNs have been observed until now (e.g. the V-C
catalog~\citep{200600000_Veron} has $694$ AGNs with redshift $z \leq
0.024$). Similarly, it is reasonable to assume that other possible
sources of UHECRs are not too far away, because UHECRs are not
isotropically distributed in the celestial sphere.

\subsubsection{Confirmation of the UHECR Sources}

The assumption in the above paragraph is that we have confirmed the
source, or we are convinced of the fact that different signals come
from the same source. However, it is not always the case. Notice the
fact that UHECRs are singular events (it seldom happens that the
UHECR events have clustering properties), confirm their sources by
statistical correlation is very important.

Metrical bias in spatial dimensions are caused by (i) intergalactic
magnetic fields and (ii) the uncertainties of detectors; the LIV
effect cannot affect the orientation of UHECRs. If the collective
effect of (i) and (ii) is small enough, we can confirm the sources
by their locations in the celestial sphere; however, it may not be
the case. If we assume that the effects of (i) and (ii) are both
stochastic, confirmation of the sources is a pure statistical
inferential problem. Astrophysical parameters only affect the
statistical samples by (i) the UHECR energy band or (ii) the
possible correlative time interval.
\cite{200700000_Cho_science_UHECR_from_galaxies} has already
discussed the statistical correlation between the arrival directions
and the positions of known AGN. The same method can be used for our
purpose; however, their arguments do not include the temporal
dimension. When discussing the LIV effect, temporal dimension is
very important. Hence, we should put by hand a possible correlative
time interval when choosing the statistical samples; that is, assume
that the collective time-delay caused by LIV, intergalactic magnetic
fields and other reasons does not exceed this interval. Notice the
fact that the observational history of UHECRs and correlative
sources are at most several decades, which may be shorter than the
collective time-delay, it is a good idea to ignore the temporal
dimension and choose all the samples we know to do the statistical
correlation. However, if the intergalactic magnetic fields are
sufficiently large, we will never know the sources of UHECRs, no
matter whether LIV exists or not.

\subsection{Problem of the Intergalactic Magnetic Fields\label{subsection_mag_field}}

The third but the most annoying problem is the intergalactic
magnetic fields. Because protons take charges, their trajectories
will be (Larmour) curved by magnetic fields, and the departures from
straight lines will cause extra time-delays. Our method is only
suitable when the time-delay $\Delta t_{\mathrm{M}}$ by the magnetic
fields is less than by the LIV effect.

Because an UHECR particle keeps constant energy inside some
homogeneous magnetic field, the time-delay should be
\begin{equation}
    \Delta t_{\mathrm{M}} \simeq \frac{1}{24} \frac{D^3}{c
    r_\mathrm{L}^2}
        \simeq 0.79 \left( \frac{D}{3\,\mathrm{kpc}} \right)^3
        \left( \frac{E}{6.3 \times 10^{19}\,\mathrm{eV}} \right)^{-2}
        \left( \frac{B_{\bot}}{1\,\mathrm{\mu G}} \right)^2\,\mathrm{yr}
        \mbox{,}
    \label{equation_const_field}
\end{equation}
where $D$ is the linear distance of the trajectory, $B_{\bot}$ is
the perpendicular magnitude of the magnetic field, $E$ is the energy
of the particle, and $r_\mathrm{L} = E / (c \cdot e B_{\bot})$ is
the Larmour radius.

\subsubsection{Comparison with Photons\label{subsubsection_time-delay_I}}

For simplicity, we first discuss the way of comparing the UHECRs'
time-delay with photons, because photons are irrespective of the
magnetic field, and their time-delay by the LIV effect can be
neglected compared to UHECRs for their relatively lower energies.

The real trajectory can be devided into three parts, inside the host
galaxy, inside our Galaxy and in the intergalactic media (IGM), that
is, $\Delta t_{\mathrm{M}} = \Delta t_{\mathrm{M,host}} + \Delta
t_{\mathrm{M,Milky}} + \Delta t_{\mathrm{M,IGM}}$. We have already
chosen the values of $D$ and $B_{\bot}$ both for a typical galaxy in
Eq.~(\ref{equation_const_field}), so $\Delta t_{\mathrm{M,Milky}}
\sim 0.79\,\mathrm{yr}$ is the typical value for the time-delay
effect of the Milky Way, which can be negligible compared to $\Delta
t_{\mathrm{QG}}$ we estimated in Eq.~(\ref{equation_an_example}). Of
course, $\Delta t_{\mathrm{QG}}$ decreases when the source comes
nearer, but the effect by the Milky Way's magnetic field remains
unaltered, so it would be troublesome when considering the use of
more nearby sources to test LIV, as mentioned in
\S~\ref{subsection_time_delay}. However, because the time-delay by
magnetic fields is absolutely classical, when we have fine structure
models for the magnetic field of our Galaxy someday, we can deduct
this effect directly~\footnote{Because the correlation length of the
magnetic field in our galaxy should be compared with the scale of
the galaxy itself. Further more, we know the direction where the
UHECR particle is related to the local structure.}. When the UHECR
particles are initially more energetic than the GZK threshold,
effect from the Milky Way's magnetic field can always be negligible,
as shown in Fig.~(\ref{pic_energy_time_delay}, bottom). The
time-delay by the host galaxy of the GRB will not be worse than by
our Galaxy, because the energy $E$ will be larger (if it formally
exceeds the GZK threshold) or at less equal (if less than the GZK
threshold) when just emitted.

However, the effect by the large from scale intergalactic magnetic
fields is more thorny, because until now we lack good models for the
magnitude and topological structure of the fields. A constraint from
the CMB anisotropy~\citep{199700000_Barrow_PhysRevLett.78.3610}
gives
\begin{equation}
    B_{\mathrm{IGM}} < 6.8 \times 10^{-9} (\Omega_{0} h^2)^{1/2}
   \,\mathrm{G} \sim 4.9 \times 10^{-9}\,\mathrm{G}
    \mbox{,}
    \label{equation_constraint_CMB}
\end{equation}
where we choose $\Omega_{0} = 1$ and Hubble constant $H_{0} = 72\,
\mathrm{km\,s^{-1}\,Mpc^{-1}}$. Another constraint from the observed
rotation measure (RM) of quasars~\citep{199400000_Kronberg_rp} gives
\begin{equation}
    B_{\mathrm{IGM}} < 10^{-9} \left( \frac{\lambda}{1\,\mathrm{Mpc}}
    \right)^{-1/2}\,\mathrm{G}
    \mbox{,}
    \label{equation_constraint_AGN}
\end{equation}
where $\lambda$ denotes the correlation length (coherence length) of
the magnetic fields, as a reasonable assumption that the power
spectrum of magnetic fields has a large scale cut-off.

\begin{figure}[ht]
\includegraphics[height=11.2cm]{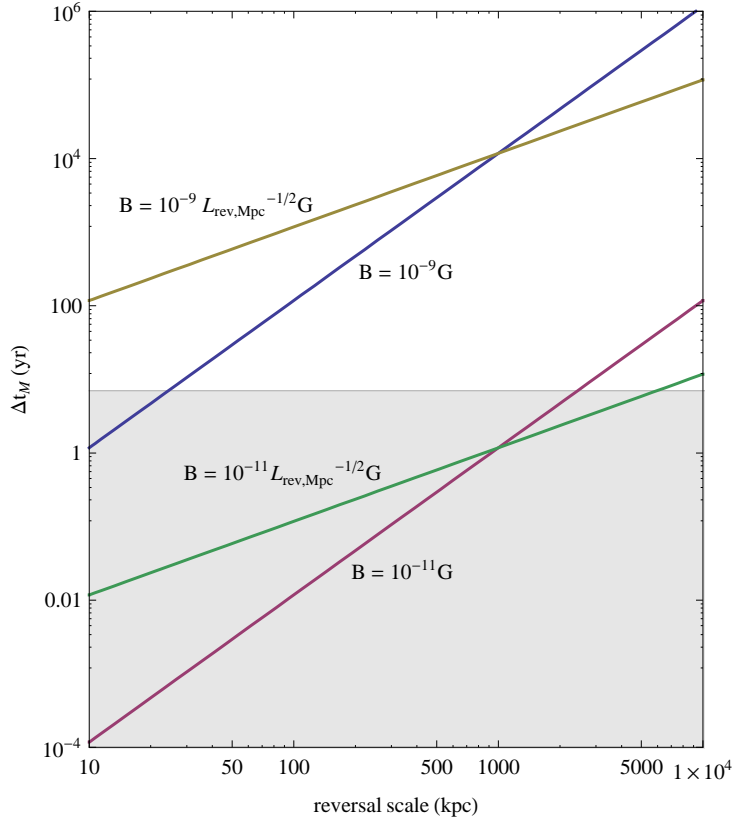}
\caption{Time-delay $\Delta t_{\mathrm{M,IGM}}$ caused by the
intergalactic magnetic fields when $z = 0.1 \simeq 400 \mathrm{Mpc}$
and $E = 6.3 \times 10^{19}\,\mathrm{eV}$, with different
correlation length and strength of the magnetic fields. The
horizontal line is $\Delta t_{\mathrm{QG}} = 7\,\mathrm{yr}$, as the
example shows in Eq.~(\ref{equation_an_example}). The method is
useful when $\Delta t_{\mathrm{M,IGM}} < \Delta t_{\mathrm{M}} <
\Delta t_{\mathrm{QG}}$.} \label{pic_delta_t_M}
\end{figure}

If we assume that the field is conglomerated and homogeneous inside
every segment (with typical scale of correlation length $\lambda$),
the UHECR particle will randomly change its direction due to
Larmour's motion, but goes a nearly strict line as a whole. If
$B_{\mathrm{IGM}}$ is independent the correlation length $\lambda$,
the overall time-delay should be
\begin{equation}
    \Delta t_{\mathrm{M,IGM}}
        \simeq 1.18
        \left( \frac{D}{400\,\mathrm{Mpc}} \right)
        \left( \frac{\lambda}{1\,\mathrm{Mpc}} \right)^2
        \left( \frac{E}{6.3 \times 10^{19}\,\mathrm{eV}} \right)^{-2}
        \left( \frac{B_{\bot}}{10^{-11}\,\mathrm{G}} \right)^2\,\mathrm{yr}
        \mbox{.}
        \label{eq_Delt_CMB}
\end{equation}
When $B$ depends on the correlation length as $B = B_0
\lambda^{-1/2}\,\mathrm{G}$ in Eq.~(\ref{equation_constraint_AGN}),
the time-delay is
\begin{equation}
    \Delta t_{\mathrm{M,IGM}}
        \simeq 1.18
        \left( \frac{D}{400\,\mathrm{Mpc}} \right)
        \left( \frac{\lambda}{1\,\mathrm{Mpc}} \right)
        \left( \frac{E}{6.3 \times 10^{19}\,\mathrm{eV}} \right)^{-2}
        \left( \frac{B_{0,\bot}}{10^{-11}\,\mathrm{G}} \right)^2\,\mathrm{yr}
        \mbox{.}
        \label{eq_Delt_RM}
\end{equation}
Other possible parameters are denoted in Fig.~(\ref{pic_delta_t_M}).
We see that it is needed~\footnote{It is possible in principle
because $B_{\mathrm{IGM}}$ remains largely unknown by the intrinsic
observational
difficulties~\citep{199600000_Beck_annurev.astro.34.1}.} for our
purpose that $B_{\mathrm{IGM}}$ is slightly less than the upper
limits given by Eq.~(\ref{equation_constraint_CMB}) and
(\ref{equation_constraint_AGN}), unless we choose a smaller
correlation length.

\begin{figure}[ht]
\includegraphics[height=9.2cm]{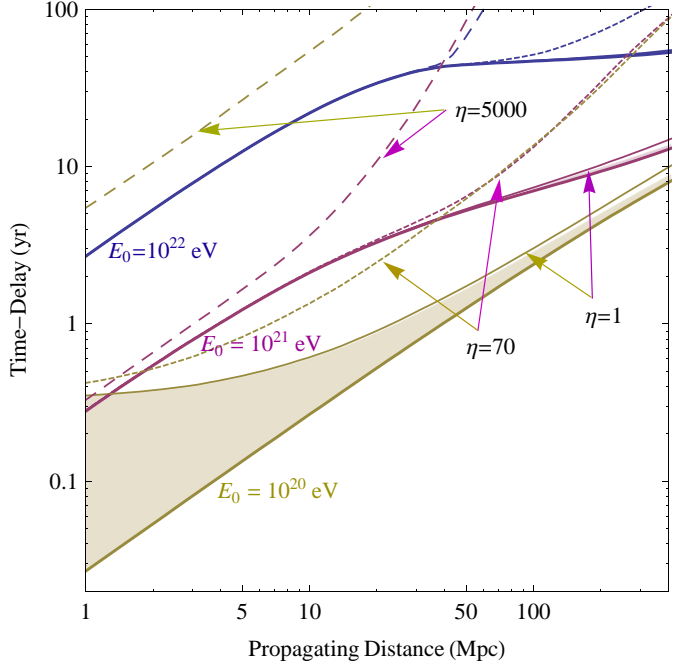}
\caption{The thick solid lines and shadow regions are the same as in
Fig.~(\ref{pic_energy_time_delay}). In addition, we calculated
$\Delta t_{\mathrm{QG}} + \Delta t_{\mathrm{M,Milky}} + \Delta
t_{\mathrm{M,IGM}}$ with $\eta = 1$, $70$ and $5000$ respectively.
It seems that finding the sources of the UHECR events with energy $E
> 3 \times 10^{20}\,\mathrm{eV}$ will tremendously help us to
(dis)confirm the LIV effect. See the context for detail.}
\label{pic_different_mag_field}
\end{figure}

Things will be more interesting when consider the UHECR particles
with energy exceeding the GZK threshold. Noting that $B_{\bot}$
($B_{0,\bot}$) and $\lambda$ in Eq.~(\ref{eq_Delt_CMB}) and
(\ref{eq_Delt_RM}) are independent of the source properties, we may
define
\begin{equation}
    \eta \equiv 1.18 \left( \frac{\lambda}{1\,\mathrm{Mpc}} \right)^2
        \left( \frac{B_{\bot}}{10^{-11}\,\mathrm{G}} \right)^2
\end{equation}
in Eq.~(\ref{eq_Delt_CMB}) and
\begin{equation}
    \eta \equiv 1.18 \times \left( \frac{\lambda}{1\,\mathrm{Mpc}} \right)
        \left( \frac{B_{0,\bot}}{10^{-11}\,\mathrm{G}} \right)^2
\end{equation}
in Eq.~(\ref{eq_Delt_RM}); then the effect by intergalactic magnetic
field has a uniform expression
\begin{equation}
    \Delta t_{\mathrm{M,IGM}}
        \simeq \eta \cdot \left( \frac{D}{400\,\mathrm{Mpc}} \right)
        \left( \frac{E}{6.3 \times 10^{19}\,\mathrm{eV}} \right)^{-2} \,\mathrm{yr}
        \mbox{.}
\end{equation}
In Fig.~(\ref{pic_different_mag_field}), we calculated $\Delta
t_{\mathrm{QG}} + \Delta t_{\mathrm{M,Milky}} + \Delta
t_{\mathrm{M,IGM}}$ in all, with $\eta = 1$, $70$ and $5000$
respectively. $\eta = 5000$ has already saturated the upper bound
given by Eq.~(\ref{equation_constraint_CMB}) and
(\ref{equation_constraint_AGN}), so $\Delta t_{\mathrm{M,IGM}}$
cannot be larger. Noting that when $E_{0} \geq
10^{21}\,\mathrm{eV}$, the UHECR particle will absolutely not be
affected by magnetic fields if it is not too far away (roughly $D
\leq 10\,\mathrm{Mpc}$), hence the \emph{only} thing that can make a
visible time-delay is the LIV effect. HiRes and AGASA have already
observed a couple of the UHECR events with energy $E > 3 \times
10^{20}\,\mathrm{eV}$~\citep{200200000_HiRes,200300000_Takeda_AstropartPhy}.
If their distance $D > 20\,\mathrm{Mpc}$, their initial energy
$E_{0}$ will exceed $10^{21}\,\mathrm{eV}$, as shown in
Fig.~(\ref{pic_energy_time_delay}, top). Hence, seeking the sources
of UHECRs with energy $E > 3 \times 10^{20}\,\mathrm{eV}$ will
tremendously help us to (dis)confirm the LIV effect.

Things will be worse if there exists a \emph{global} cosmic magnetic
field, or the fields have structures like filaments or
sheets~\citep{199800000_Ryu}. Those will cause larger time-delays.
The first trouble can be easily seen from the expression of $\Delta
t_{\mathrm{M,IGM}}$ given above, because it is equivalent to a huge
$\lambda$. The second trouble is because, when a magnetic cloud
collapses to $1/k$ of its diameter, the magnetic field strength will
increase $k^2$ times than it used to be. Although the particle will
miss a lot of clouds it used to bang into (only for the case of
filaments but not sheets, which make things worse), we still have
\begin{equation}
    B_{\bot}^2 \lambda^2 \rightarrow \frac{1}{k}
    (k^2 B_{\bot})^2 \left(\frac{1}{k} \lambda\right)^2 = k
    B_{\bot}^2 \lambda^2
    \mbox{,}
\end{equation}
and the same for the $B_{\bot}^2 \lambda$ cases.

However, in fact the irrefutable observed anisotropy of
UHECRs~\citep{200700000_Cho_science_UHECR_from_galaxies} has already
given us an upper limit for $B_{\bot}$ and $\lambda$, and also
whether the magnetic field has already collapsed to filaments or
sheets or not. Note that 20 among 28 highest energy events detected
by the Pierre Auger Observatory are within a $3.1^{\circ}$ circle of
nearby AGNs (with distance less than $75\,\mathrm{Mpc}$ away). No
matter whether we believe these UHECR particles to originate from
those AGNs or not, it is unassailable that UHECRs are
unisotropically distributed, and seem to be correlated to the
Super-galactic plane. So, the angular dispersion caused by the
intergalactic magnetic fields should be less than a couple of
degrees. The angle departure inside some homogeneous field is
\begin{equation}
    \alpha \simeq \frac{D}{2 r_{\mathrm{L}}} \mbox{,}
\end{equation}
and the different irrelevant magnetic bulks (with typical size
$\lambda$) can be considered as a random walking process. The
overall angle departure is
\begin{equation}
    \alpha \simeq \frac{\lambda}{2 r_{\mathrm{L}}}
    \sqrt{\frac{D}{\lambda}}
        = \frac{\sqrt{D \cdot \lambda}}{2 r_{\mathrm{L}}} \mbox{.}
\end{equation}
Choosing $D = 100\,\mathrm{Mpc}$ as the typical scale of
Super-galactic plane, we have
\begin{equation}
    \alpha_{\mathrm{Milky}} \simeq 1.26^{\circ}
        \left( \frac{D}{3\,\mathrm{kpc}} \right)
        \left( \frac{E}{6.3 \times 10^{19}\,\mathrm{eV}} \right)^{-1}
        \left( \frac{B_{\bot}}{10^{-6}\,\mathrm{G}} \right)
        \mbox{,}
\end{equation}
and
\begin{equation}
    \alpha_{\mathrm{IGM}} \simeq 4.21^{\circ}
        \left( \frac{D}{100\,\mathrm{Mpc}} \right)^{1/2}
        \left( \frac{\lambda}{1\,\mathrm{Mpc}} \right)^{1/2}
        \left( \frac{E}{6.3 \times 10^{19}\,\mathrm{eV}} \right)^{-1}
        \left( \frac{B_{\bot}}{10^{-9}\,\mathrm{G}} \right)
        \mbox{.}
\end{equation}
Notice that when $\alpha_{\mathrm{IGM}}$ approaches a couple of
degrees, as the above equation shows, the upper limits given by
Eq.~(\ref{equation_constraint_CMB}) and
(\ref{equation_constraint_AGN}) have already been saturated. In
addition, because of the fact that $\lambda^{1/2} B_{\bot}
\rightarrow \sqrt{k} \lambda^{1/2} B_{\bot}$, filaments or sheets
can also be suppressed.

One question is whether the Pierre Auger data tell us that
$\alpha_{\mathrm{IGM}}$ should be equal to (rather than less than)
several degrees? Absolutely not. $\alpha_{\mathrm{IGM}}$ can also be
much less (so $B$ and $\lambda$ can also be much less). Even if we
have confirmed some UHECR sources, the angular dispersion can also
be caused by reasons other than $\alpha_{\mathrm{IGM}}$, for
example, magnetic field of the Milky way or simply the measurement
errors.

We notice that some authors gave a larger $\Delta
t_{\mathrm{M,IGM}}$ compared to ours given in
Eq.~(\ref{equation_constraint_CMB}) and
(\ref{equation_constraint_AGN}). \cite{199600000_waxman_blurred}
gave $\Delta t_{\mathrm{M,IGM}} \sim 100\,\mathrm{yr}$ because their
correlation length $\lambda \sim 10\,\mathrm{Mpc}$ is 10 times
larger than ours (equivalent to $\eta = 100$ in our definition).
\cite{200100000_Sigl_UHECR} gave $\Delta t_{\mathrm{M,IGM}} \sim
10^3\,\mathrm{yr}$ because he chose a really large magnetic field $B
\sim 10^9\,\mathrm{G}$ (equivalent to $\eta = 10^4$); however, with
a smaller travelling distance $D$. \cite{200100000_waxman_UHECR}
gave an \emph{upper} bound of $\Delta t_{\mathrm{M,IGM}}$ even as
large as $10^7\,\mathrm{yr}$, because his typical magnetic field $B
\sim 10^8\,\mathrm{G}$ is really huge. He also argued that $\Delta
t_{\mathrm{M,IGM}} > 100\,\mathrm{yr}$ by some statistical reasons
of nearby source candidates and the UHECR events above the GZK
threshold. The first two estimations are consistent with our
constraint from correlation of Super-galactic plane and the UHECR
events, the few discrepancies are just because we choose different
typical parameters (which are all possible according to our current
knowledge, because we know really little about the true value of $B$
and $\lambda$) to write our formulas. We suggest that the anisotropy
of UHECRs can give a tighter constraint of intergalactic magnetic
field strength $B$, so the upper bound of $\Delta
t_{\mathrm{M,IGM}}$ we can assure at present should be as low as
$10^{4-5}\,\mathrm{yr}$. The \emph{lower} bound $\Delta
t_{\mathrm{M,IGM}} > 100\,\mathrm{yr}$ can be overcome because we
know really little about both possible nearby sources and the UHECRs
events, and the estimation is dependent on some details of source
models. In addition, all the estimations given above are only
suitable for particles with energy below the GZK threshold, because
the energy loss is ignored. As we show in
Fig.~(\ref{pic_different_mag_field}), the effect of intergalactic
magnetic field is much less important if the energy of the
\emph{observed} UHECR event is much larger than the GZK threshold.

\subsubsection{Comparison with Other UHECR Events\label{subsubsection_time-delay_II}}

We can also compare the time-delay with other UHECR events (with
slightly different energies), emitted nearly simultaneously from the
same source. Of course, because the UHECR events are really rare, it
may hardly happen.

In this case, blurs in both arrival direction and time-delay have to
be analyzed carefully. (i) Blurs have two reasons. Particles with
different energy follow different trajectories, thus leading to
different directions and time-delays, because of the random
topological distributions of the intergalactic magnetic fields. (ii)
At the same time, particles above the GZK threshold energy would
interact with CMB photons as the Poisson processes, introducing
extra randomicity. \cite{199600000_waxman_blurred} discussed the
blur effect with UHECRs below the GZK threshold, in which case
energy loss by photomeson production can be ignored. At the end of
\S~\ref{subsection_time_delay}, we have already discussed a little
the influence of the LIV time-delay by stochastic photomeson
production.

It is easy to understand that when the particles are extremely
energetic, blurs in both arrival direction and time-delay caused by
the intergalactic magnetic fields become less important. However,
using one of the UHECRs to scale the others may be dangerous, if
their energies are large enough (e.g. larger or equal to the GZK
threshold) to make us believe that they have suffered one or more
times the photomeson interactions. Because of the randomicity of
Poisson arrival photomeson interactions, particles observed with the
same energy from the same source may have tremendously different
interacting histories and thus possess different time-delays caused
by both the intergalactic magnetic fields and the LIV effects.

However, for UHECRs less energetic than the GZK threshold,
photomeson production is turned off, and comparison becomes
possible. The requirement that the intergalactic magnetic fields
should not be very large, is the same as in the case of comparing
UHECRs with photons, which we have already discussed in
\S~\ref{subsubsection_time-delay_I}.

\subsection{Problem of the Energy Measurements in Air Shower Detectors}

Notice that the energy measurements in different mass composition of
the UHECR events and different air shower detectors have
disagreements from each other which cannot be negligible, so it is
necessary to discuss here the influence of the LIV confirmation by
energy demarcation uncertainties. In \S~\ref{subsection_mag_field},
we have already discussed two different methods to restrict LIV, the
comparison (i) with photons and (ii) with the different UHECRs
respectively from the same source.

For the reason that the investigations of the LIV confirmation are
qualitatively rather than quantitatively at present, the energy
metrical uncertainties are not crucial for method (i), because it
can only introduce an order one coefficient of $\xi$ in
Eq.~(\ref{eq_time_delay}). When the UHECR events are not too
energetic to neglect the time-delay caused by the intergalactic
magnetic fields, \emph{absolute} energy measurements are important.
However, a global constraint for the collective influence of $\Delta
t_{\mathrm{QG}} + \Delta t_{\mathrm{M,IGM}}$, hence the upper limits
for both $\Delta t_{\mathrm{QG}}$ and $\Delta t_{\mathrm{M,IGM}}$
respectively, are still suitable for our purpose.

For method (ii), things are a little more complicated. Uncertainties
introduced by the different assumptions of mass composition are not
crucial. The reason is that, when assuming different UHECRs to be
the same kind of particles (protons in our context), a mistaken
assumptive mass composition can only introduce an order one
coefficient of $\xi$, just as in the case of method (i). However, it
is intractable for UHECRs detected by different air shower detectors
with different energy metrical techniques. A wiser way is to choose
some kind of calorimetric measurements to determine UHECRs' energies
by different detectors (like fluorescence light emissions
\citep{198300000_Linsley,199900000_Song}) which are relatively model
independent. As a matter of fact the discussions of LIV are
presently still superficial, we may hope that energy demarcations
are finer for further investigations of LIV in the near future.

\section{Discussion}

\subsection{Two Known Events}

There was an archaeological report about the association of UHECRs
and GRBs~\citep{199505009_Milgrom_ApJ1995}. The authors found that
GRB 910503 and 921230 are associated with two highest-energy
cosmic-ray shower events, with really small error boxes and
time-delays of 5.5 and 11 months respectively. If GRBs are really
sources for those two UHECR events, there are very strong
constraints both for LIV and the strength of intergalactic magnetic
fields (as the time-delay is much shorter than the naive estimation
we make in Eq.~(\ref{equation_an_example})), because all effects
such as rest mass, magnetic fields and quantum gravity, are addible,
and to ignore some of them gives the upper constraint for the rest
ones. However, we should not be too serious for that kind of
stories, because they may only be a coincidence.

\subsection{Comparison with Other Models}

Although there are other constraints of LIV which are much stronger
than the method we suggested, the method mentioned above has also
its special purpose.
Birefringence~\citep{200100000_PhysRevD.64.083007_Gleiser,200300000_Mitrofanov_nature_426139a_polarization,200702006v1_Fan_GRB_polarization__polarimetry_and_QG}
can only be calculated in the framework of loop quantum gravity but
not in other approaches, hence it may be wrong in a whole. The
synchrotron radiative
constraint~\citep{200200000_PhysRevD.66.081302_Jacobson,200300000_Jacobson_nature01882}
depends on a special
theory~\citep{200300000_PhysRevLett.90.211601_Myers}, which needs a
dimension-5 Lorentz-violating terms to induce birefringent photons
but subluminal electrons (whose maximum speed cannot converge to
$c$). The inverse proportion of the GZK anomaly may also give some
stronger constraint. However, the scattering dynamical discussions
are always only one-sided, which means that a scattering channel is
open or suppressed only if the effect of LIV in opposite for two
relative particles (and therefore their velocity as well as their
effective mass being different). Although in the old days, the GZK
anomaly is the most important reason for theoreticians to study LIV,
its
inexistence~\citep{200703099v1,200700000_Cho_science_UHECR_from_galaxies}
has not borne down the LIV subjects.

Testing the time-delay of UHECRs is a more direct way to study LIV.
It can contain most kinds of theoretical works. If the intergalactic
magnetic fields are sufficiently small (which is still absolutely
consistent with the observations until now), it may have larger
examinable parameter space for violation scale $\xi$ (in
Eq.~(\ref{eq_time_delay})) than using photons or neutrinos. Even if
its examinable parameter space is in fact much smaller, for the
reasons mentioned above, the other causations are all classical and
thus can someday be subtracted by models.

\section{Conclusions}

We have suggested to (dis)confirm LIV by simply detecting the
time-delay of UHECRs. We considered some other reasons which also
cause the time-delay, including the intergalactic magnetic fields.
If the energy of the UHECR events we observed is \emph{below} the
GZK threshold $6.3 \times 10^{19}\,\mathrm{eV}$, a typical
intergalactic magnetic field $B \lesssim 10^{-11}\,\mathrm{G}$ and
correlation length $\lambda \lesssim 1\,\mathrm{Mpc}$ may be needed
to give a parameter space examinable enough to constrain LIV.
However, for an UHECR event with energy larger than $3 \times
10^{20}\,\mathrm{eV}$, our method is always possible. Because of the
fact that we know really little about the intergalactic magnetic
field's strength, if it is much smaller than the current upper limit
$B \lesssim 10^{-9}\,\mathrm{G}$, our method may give a larger
examinable parameter space and a stronger constraint of LIV than
other constraints.

\section*{Acknowledgements}

We would like to thank Yi-Zhong Fan and Xiang-Yu Wang for helpful
discussions. This work is supported by the National Natural Science
Foundation of China (grants 10221001 and 10640420144) and the
National Basic Research Program of China (973 program) No.
2007CB815404.

\bibliographystyle{oxo_fianl_ref_style}
\bibliography{oxo_final_ref}

\end{document}